\documentclass[twocolumn,showpacs,preprintnumbers,amsmath,amssymb]{revtex4}
\usepackage{graphicx,color}
\usepackage{bm}%
\begin{document}
\title{High-frequency vibrational density of states of a disordered solid}
\author{C. Tomaras$^{1,2}$ and W. Schirmacher$^{1,2,3}$}
 \affiliation{$^1$Institut f\"ur Physik, Universit\"at Mainz, Staudinger Weg 7,
D-55099 Mainz, Germany}
 \affiliation{$^2$Inst. f. funktionelle Materialien, Phys. Dept. E13, TU M\"unchen, D-85747 Garching, Germany}
 \affiliation{$^3$Dipartimento di Fisica, Universit\'a di Roma
``La Sapienza'', P'le Aldo Moro 2, I-00185, Roma, Italy}
\begin{abstract}
We investigate the high-frequency behavior of the density of vibrational
states in three-dimensional elasticity theory with spatially
fluctuating elastic moduli. At frequencies well above the mobility
edge, instanton solutions yield an exponentially decaying density
of states. The instanton solutions describe excitations, which become
localized due to the disorder-induced fluctuations, which lower the
sound velocity in a finite region compared to its average value. The
exponentially decaying density of states
(known in electronic systems as the Lifshitz tail)
is governed by the statistics of a fluctuating-elasticity landscape,
capable of trapping the vibrational excitations. 

\end{abstract}
\maketitle

\section{Introduction}

The density of states (DOS)
of a disordered system is a quantity which has
been vividly discussed until today 
\cite{velicky69,elliott74,yonezawa75,vollhardt10}.
Lifshitz has been one of
the first authors who calculated the disorder-induced corrections
on the DOS, for electron and phonon systems \cite{lifshitz1964energy}.
By a phenomenological argument, he explained the occurrence of
an exponential tail of the DOS in the vicinity of the
band-edges, which is related to the localization of the one-particle
states. As the penetrability of an arbitrarily shaped barrier tends
towards zero at large energies \cite{landau67}, one expects
the high-energy eigenstates of a system with a random
fluctuating potential to be localized. The DOS is then
proportional to the probability of the occurance of wells, capable
of trapping the single-particle states, which for weak fluctuations
(e.g. low impurity concentrations ) can always be approximated by
an exponential function. The energy dependence of the exponent
depends crucially on the energy of localization, from which some
substantial correction from the set of maximaly crossed diagrams in
the perturbation expansion arises, similar to the energy-momentum relation
of a wave in a potential well, if the length scale set by the disorder
prameter is of the order of the localization length. Such corrections
can be treated in a field-theoretical approach by means of an $\epsilon$-expansion
of the nonlinear $\sigma$ model in the vicinity of the localization
energy \cite{hikami1981anderson,mckane1981localization}, or by self-consistent
mode coupling expansion techniques \cite{vollhardt1980diagrammatic},
which lead to the potential-well analogy \cite{economou1985quantitative,economou1983connection,economou1984localized,soukoulis1984exponential}.
These corrections are important for the electronic problem, because
the Lifshitz tail mostly consists of the ground states of such wells,
for which the irreducible 1-particle self-energy is energy dependent. 

For phonons the situation is different, because one deals with fluctuating
elastic constants
instead of potentials, and localized states appear
only at the upper band edge \cite{john83,schirmacher1998harmonic} for
positive values of $\omega^2$ 
\cite{pinski12,pinski12a}, 
where $\omega$ is the vibrational
excitation frequency.
There is an extended amount of literature concerning the cross-over from
the Debye type acoustic wave regime to a regime
of diffusive, random-matrix type of vibrational excitations
(``boson peak'' \cite{elliott84,horbach01,nakayama02}), which
has been accurately described within the $\sigma$-model approach
by two non-crossing techniques, namely the self-consistent Born
\cite{schirmacher2006thermal,schirmacher2007acoustic,schirmacher2008vibrational,
marruzzo_forthcoming2} and coherent-potential
approximation \cite{schirmacher1998harmonic,marruzzo_forthcoming1,schirm13,koehler13}.
Instead here
we are interested in the behavior at large positive frequencies.
In the $\sigma$-model approach one finds that the one-particle self-energy
on the localized side becomes frequency independent, and the system
can be assumed to be described by a renormalized Ginzburg-Landau theory
\cite{john1984electronic,cardy1978electron}. The states with the
highest values of $\omega^2$
within a region of constant elastic modulus, bounded
by a mismatch, carry wave numbers of the order of the Debye wavenumber.
Hence these states dominate the Lifshitz tail. If one assumes further
the fluctuations of the elastic constants to be small and the localization
length to scale with the inverse frequency $\xi\propto\omega^{-1}$
as in elementary wave mechanics \cite{landau67},
the DOS should be proportional to $\exp(-a\omega^{4-d})$.
Such a behavior
of the exponent is suggested by recent experiments
\cite{chumakov04,walterfang05}, in which
a linear exponential decay of the DOS in the high-frequency region is found.

In the remainder of this paper we will reformulate the field theoretic
approach within the Keldysh formulation of quantum dynamics, in order
to include the instanton contribution, from wich the tail of the
DOS
can be extracted.

\section{Keldysh formulation of weakly disordered phonons}

The quantum dynamics for the displacement-field ${\bf u}({\bf x},t)$
with spatially dependent elastic moduli will be calculated from the
Keldysh partition function \mbox{$1=\int\mathcal{D}{\bf u}e^{iS}$,} where
the action involves the classical (symmetric) and quantum (antisymmetric)
linear combinations of the two replicas acting left and right onto
the density matrix describing the inital state \cite{levchenko2007keldysh}.
The reason for naming them classical and quantum is as follows: if
the action is related to the Lagrangian of a simple non-relativistic
one-particle system, the saddle-point solution of the field theory
with \mbox{$u_{q}=u_{+}-u_{-}=0$,} \mbox{$u_{cl}\equiv u_{+}+u_{-}=u(t)$} yields
Newton's equation of motion for the particle's trajectory $u(t)$.
Quantum corrections like the appearance of a
phase or tunneling contributions arise from
finite expectation values and higher correlations of the antisymmetric
$u_{+}-u_{-}$ linear combination, which is hence named the quantum
component. 

The action may involve arbitrary nonlinear terms, which in a classic
kinetic approximation gives rise to phonon thermalization, at least
in three dimensions. For the vibrational spectrum of disordered systems
the anharmonic interactions are only important below the
boson peak \cite{tomaras10,ferrante13}.
For our
purpose it is hence enough to approximate the Keldysh action to quadratic
order in the displacement field, 
\begin{gather}
S=\int d^{d+1}x\left\{ -{\bf u}^{q}\partial_{t}^{2}{\bf u}^{cl}+2\mu(\mathbf{x})u_{ij}^{q}u_{ji}^{cl}+\lambda({\bf x})u_{ii}^{q}u_{ii}^{cl}\right\} \nonumber \\
+\int d^{d+1}x{\bf u}^{q}\left(G^{-1}\right)_{K}{\bf u}^{q}\label{eq:1}
\end{gather}

where $x=({\bf x},t)$ and $u_{ij}=\frac{1}{2}\left(\partial_{i}u_{j}+\partial_{j}u_{i}\right)$
is the usual strain-tensor, and $\lambda$ and $\mu$ are Lam\'e's elastic
constants, which are assumed to fluctuate due to the structural disorder of the material
\cite{schirmacher2006thermal,schirmacher2007acoustic,schirmacher2008vibrational,
marruzzo_forthcoming2,marruzzo_forthcoming1}.
We use units in which the mass density equals unity.
The Keldysh component $G_{K}$ characterises
the actual state of the system, which can be determined from knowledge
of the retarded and advanced Green's function and a proper initial
condition. If one assumes that the disorder does not alter the thermalisation
process, the Keldysh component can safely be replaced by a thermal
distribution. The formulation (\ref{eq:1}) ignores further initial
correlations. 

The advantage of Keldysh's closed time-contour is the absence of vacuum
contributions to the partition function, hence one can avarage it
directly over an arbitrary distribution of elastic constants. This
average can be performed formally by characterizing the probability
distribution through its irreducible correlation functions 
\begin{equation}
K(x_{1},...,x_{n})\equiv\dfrac{\delta}{\delta x_{1}}....\dfrac{\delta}{\delta x_{n}}\ln\int d\mu P[\mu]e^{(\mu,j)}\, ,
\end{equation}
which leads to an action of the form

\begin{gather}
S_{dis}=S-S[\langle\mu\rangle,\langle\lambda\rangle]=\nonumber \\
+\int_{x_{1}x_{2}}K_{\mu}({\bf x}_{1},{\bf x}_{2})u_{ij}^{q}(x_{1})u_{ji}^{cl}(x_{1})u_{lm}^{q}(x_{2})u_{ml}^{cl}(x_{2})\nonumber \\
+\int_{x_{1}x_{2}}K_{\lambda}({\bf x}_{1},{\bf x}_{2})u_{ii}^{q}(x_{1})u_{ii}^{cl}(x_{1})u_{jj}^{q}(x_{2})u_{jj}^{cl}(x_{2})+...\,.\label{eq:2}
\end{gather}
 For the Gaussian theory the structure of the action is the same as
for a dissipative quantum system \cite{kamenev2009keldysh}, in which
the spectral function of the heat bath is replaced by two classical
strain fields. Therefore the calculation of instanton solutions in
disordered systems exhibit a strong resemblance to the calculation
of thermal activation of a system attached to some heat bath \cite{kamenev2009keldysh}. 

Within this article we do not want to perform technically 
detailed calculations.
We rather want to demonstrate the general procedure of mapping
the disordered bosonic system onto
a disordered fermionic one within the Keldysh prescription. Of
course this mapping is only possible between the pair modes of the
bosonic and fermionic systems. 

In general the theory can be represented in terms of these 2-point
functions using the Fadeev-Popov transformation
\begin{gather*}
F\left[u_{ij}^{\alpha}(x_{1})u_{lm}^{\beta}(x_{2})\right]=\int_{Q,\Lambda}e^{i\mathrm{Tr}\left[\left(\Lambda-uu^{\dagger}\right)Q\right]}\\
\equiv\int\mathcal{D}Q\mathcal{D}\Lambda F[Q_{ijlm}^{\alpha\beta}(x_{1},x_{2})]e^{i\Lambda_{ijlm}^{\alpha\beta}\left(Q_{mlji}^{\beta\alpha}-u_{ml}^{\beta}(x_{1})u_{ji}^{\alpha}(x_{2})\right)}.
\end{gather*}
Here Greek indices represent the Keldysh degrees of freedom. The following
mapping achieves the same causality structure for the composite phonons
as for the electrons \cite{kamenev2009keldysh}
\begin{gather*}
A_{R}\equiv A\left(\begin{array}{cc}
0 & 1\\
1 & 0
\end{array}\right),A_{L}\equiv\left(\begin{array}{cc}
0 & 1\\
1 & 0
\end{array}\right)A\,.
\end{gather*}
 If the matrix $A$ has the causality structure of a bosonic Green's
function, $A_{R}$ has the c. s. of a fermionic Green's function and
$A_{L}$ the c.s. of a fermionic self energy. The advantage is now,
that the phonon nonlinear $\sigma$-model can be developed along
the eletronic counterpart. One can now write the disorder-induced
nonlinear terms as 
\begin{gather}
S_{dis}=\sum_{n}\mathrm{Tr}\left[K_{n}^{\mu}Q_{R}^{n}+K_{n}^{\lambda}\left(\chi Q_{R}\right)^{n}\right]\label{finalaction}
\end{gather}
in which $\chi$ is a characteristic matrix defined according to $\chi Q_{ijlm}=\delta_{ij}\delta_{lm}Q_{ijlm}$. 

For sake of simplicity we will only consider a locally fluctuating shear
modulus , 
\mbox{$K_{n}^{\lambda}=0$},
\mbox{$K_{2}^{\mu}=\gamma^{-1}\delta({\bf x}-{\bf y})$},
\mbox{$K_{n\neq 2}^{\mu}=0$.}

In absence of anharmonicities one can immediatly integrate out the
composite field $\Lambda$, which reduces the Fadeev-Popov procedure to the
Hubbard Stratonovich transformation. This could be the starting point
to formulate the nonlinear $\sigma$ model approach on phonon localization
around the weakly disordered SCBA saddle point \cite{schirmacher2006thermal}. 

However, as also discussed by Cardy \cite{cardy1978electron}, in
the large energy regime, where the phonon energy is large compared
to the energy-fluctuations set by the disorder potential, the semiclassical
one-particle approximation of the partition function becomes valid.
It is more convenient to discuss the Lifshitz-tail in terms of the
one-particle functions, as one avoids the complicated 
renormalization-group method. One
would agree, that the simple one-particle saddle point is sufficient,
as long as one is interested in the deep localization regime, where
knowledge of the mobility edge is lost. In contrast, the universal
properties of vibrations in a glass in the vicinity of the mobility edge
\cite{pinski12,pinski12a}, 
must be developed from (\ref{finalaction}) within the usual Keldysh nonlinear
$\sigma$ model.

\section{Instanton solutions}

In the remainder we will use the simple instanton picture in order
to calculate the exponential dependence of the vibrational density
of states at high energies. 

The first step is to express the partition function into an infinte
product of discrete frequencies $1=\prod_{\omega}Z(\omega)$, where
$Z(\omega)=e^{is(\omega)}$

and 
\begin{gather}
s(\omega)=\int d^{d}x\left[\omega^{2}\vec{u}_{q}\vec{u}_{cl}+g_{k}\vec{u}_{q}\vec{u}_{q}+2\mu u_{ij}^{q}u_{ji}^{cl}+\lambda u_{ii}^{q}u_{ii}^{cl}\right]\label{eq:4}\\
+\int d^{d}x\gamma\left(u_{ij}^{q}(-\omega,x)u_{ji}^{cl}(\omega,x)\right)^{2}\nonumber 
\end{gather}

If the frequency is sufficiently large, this field theory can be solved
in the semiclassical saddle-point approximation by extremizing the
action with respect to the classical and quantum components. In analogy
to the theory of dissipative quantum systems there exist instanton
solutions in which the expectation value of the quantum component
$\langle u_{i}\rangle=iv_{i}^{q}$ is finite and purely imaginary. 

From the equation of motion (\ref{eq:5}) one deduces that these solutions
describe a situation, in which condensation of strain $iv_{ij}^{q}=-v_{ij}=-v_{ij}^{cl}$
leads to a finite region with a sound velocity substancially lower
than the average. The states are hence trapped in this finite region.
There is a second instanton equation describing the other possibility,
where the sound velocity is raised. We seek for an instanton ansatz
where the saddle-point solution satisfies 
\begin{gather}
(\omega^{2}-(\lambda+\mu[v])\nabla\nabla\circ-\mu[v]\Delta)\vec{v}(\omega,x)=0\label{eq:5}\\
\mu[v]=\mu(1-\gamma v_{ij}v_{ji})\,.
\nonumber
\end{gather}

Reinsertion of this equation of motion (\ref{eq:5}) into (\ref{eq:4})
yields a finite action $if(\omega)$, and hence an exponential factor
of the partition function $e^{-f(\omega)}$, which is calculated via
formula (\ref{eq:10}).

In order to calculate the density of states and the Green's function
we have to expand (\ref{eq:4}) with respect to the fluctuations above
this instanton saddle-point $u_{ij}^{\alpha}=v_{ij}+\delta u_{ij}^{\alpha}$.
The irreducible retarded Green's function is by definition just the
correlator of $\langle\delta u_{i}^{cl}\delta_{j}u^{q}\rangle.$ Obviously
this quantity is proportional to the factor $e^{-f(\omega)}$ which
is nothing than the exponential Lifshitz-tail. The action (\ref{eq:4})
expanded among the instanton saddlepoint reads 

\begin{gather}
s(\omega)=if(\omega)+O[\delta u_{ij}^{3},\delta u_{ij}^{4}]+\int d^{d}x\omega^{2}\delta\vec{u}_{q}\delta\vec{u}_{cl}\\
+\int d^{d}x\left(2\mu(1-2\frac{\gamma}{\mu})\delta u_{ij}^{q}v_{hk}v_{kh}\delta u_{ji}^{cl}+\lambda\delta u_{ii}^{q}\delta u_{ii}^{cl}\right)
\end{gather}

The next step is to use the Bosonization-procedure (\ref{eq:4}) to
express this action in terms of the pair modes $Q_{R}=\langle\delta u_{i}^{cl}\delta u_{j}^{q}\rangle.$
Using a further saddle-point approximation in order to determine $Q_{R}$
yields to the following set of equations 
\begin{gather}
\langle Q_{R}(\omega,x)\rangle\nonumber \\
\equiv i\gamma e^{-f}\left(\frac{1}{G_{0}^{-1}(\omega,x)-\nabla\left(\Sigma+v_{ij}v_{ji}\right)_{\omega,x}\nabla^{T}}+\vec{v}\vec{v}^{T}\right)\label{eq:7}\\
\Sigma_{R}(\omega,x)\propto e^{-f}\gamma\mathrm{Tr}\nabla\frac{1}{G_{0}^{-1}(\omega,x)-\nabla\left(\Sigma+v_{ij}v_{ji}\right)_{\omega,x}\nabla^{T}}\nabla^{T}\label{eq:8}\\
f(\omega)=\int_{{\bf x}}\left\{ \omega^{2}|\vec{v}(\omega,x)|^{2}+2\mu v_{ij}v_{ji}+\lambda v_{ii}^{2}\right\} \label{eq:9}
\end{gather}

in which the fields satisfies the exact equation of motion 

\begin{gather}
\dfrac{\delta}{\delta v^{\alpha}}s(\omega)=\dfrac{\delta}{\delta v^{\alpha}}\left\{ \vec{v}^{T}(G_{0}^{-1}+\nabla\Sigma\nabla^{T})\vec{v}+S_{dis}[v]\right\} =0\,.\label{eq:10}
\end{gather}

These equations have a rather simple interepretation; they allow for
a self-consistent determination of the instanton solution in Hartree-approximation
and a further non-crossing approximation of the self-energy of the
propagator, which has a frequency and spatially dependent sound-velocity,
due to the finite instanton amplitude. However, we are mostly interested
in a discussion of the exponential factor and leave the numerical
solution of this set of equations for future work. 

From (\ref{eq:7}) it becomes clear that within this approximation
the density of states is just the usual one in SCBA approximation,
multiplied with the instanton factor $e^{-f(\omega)}$. In order to
estimate the power-law satisfied by the exponent of the DOS we look
for a spherically symmetric solution with longitudinal polarisation
in 3 dimensions. 

The equation for the radial part reads 

\begin{gather}
\omega^{2}\Psi(r)+(1-4\gamma\left(\partial_{r}\Psi\right)^{2}))\Delta_{r}\Psi(r)=0\label{eq:11}
\end{gather}

The frequency and the disorder parameter can immediatly be scaled
out according to $\phi=\dfrac{1}{\omega\sqrt{\gamma}}\Psi(\omega r)$,
where the spatially dependent part $\Psi(y=\omega r)$ satisfies the
reduced equation
\begin{gather}
\Psi(y)+(1-4\left(\partial_{y}\Psi\right)^{2}))\Delta_{y}\Psi(y)=0\label{eq:12}
\end{gather}
For numerical solution one replaces (\ref{eq:12}) with the first-order
system 

\begin{gather*}
\partial_{r}\Psi(r)=\phi(r)\\
\partial_{r}\phi(r)=\Psi(1-4\phi(r)^{2})^{-1}-2r^{-1}\phi(r)
\end{gather*}
 which may readily be solved by means of a second-order Runge Kutta
alogrithm. 

The phase can be estimated at large frequencies

\begin{gather}
f=\int d^{d}r\phi(r)(\omega^{2}+\Delta_{r}^{2})\phi(r)\nonumber \\
=\dfrac{1}{\gamma}\omega^{-d}\int d^{d}y\Psi(y)(1+\Delta_{y})\Psi(y)\nonumber \\
=\int_{q<q_{D}\omega}\frac{d^{d}q}{(2\pi)^{3}}\Psi(q)(1+q^{2})\Psi(q)\nonumber \\
\omega\rightarrow\infty:\,\frac{1}{\gamma}\left(\dfrac{\omega}{\omega_{D}}\right)^{4-d}\int\frac{d^{d}q}{(2\pi)^{3}}\Psi(q)\Psi(q)=\frac{1}{\gamma}\left(\dfrac{\omega}{\omega_{D}}\right)^{4-d}
\end{gather}

From (14) it is clear, that there exists a localized exponential solution,
as long as the state has a single point where the condensed dimensionless
strain exceeds the critical value $\partial_{y}\Psi(y)=\frac{1}{2}$,
which can always be imposed as a boundary condition. Note that $q$
is not the Fourier component of the spatial variable $r$, but of the
frequency scaled variable $y=\omega r$. 

In contrast to the electronic calculation the phase is dominated by
the small-distance behavior of the wave function. As a result we
find the same power law of the exponent as predicted by the Lifshitz-argument.

\section{Summary and Conclusion}

In this work, we formulated the theory of weakly fluctuating elastic
constants of a glass within the Keldysh-formalism, in order to propose
an expression for the high-frequency behavior of the density of states
of a glass. In three dimensions, where this approach is valid, it
yields an exponential decaying density of states. This exponential
decay is expected by basic Lifshitz-like considerations,
and has also been observed in recent experiments \cite{chumakov04,walterfang05}.
If one
measures this exponential decay in an experiment, it is therefore
straightforward to extract the disorder parameter $\gamma$, which
can be compared with the values extracted from experiments
and simulations for describing the boson peak
\cite{schirmacher2006thermal,schirmacher2007acoustic,schirmacher2008vibrational,
marruzzo_forthcoming2,marruzzo_forthcoming1}.

\end{document}